\documentclass[prl, reprint, twoside, superscriptaddress, amsmath, amssymb, aps]{revtex4-2}
\usepackage{CJKutf8}
\usepackage{amsmath}
\usepackage{enumerate}
\usepackage{url}
\usepackage{float}
\usepackage{booktabs}
\usepackage{physics}
\usepackage{amssymb}
\usepackage{graphicx}
\usepackage{placeins}
\usepackage[english]{babel}
\usepackage{booktabs}
\usepackage{makecell}
\usepackage{bm}
\usepackage{relsize}
\usepackage[bb=boondox]{mathalfa}
\DeclareMathAlphabet{\mymathbb}{U}{bbold}{m}{n}
\usepackage{hyperref}
\usepackage{physics}

\hypersetup{hypertex=true,
colorlinks=true,
linkcolor=blue,
anchorcolor=blue,
citecolor=blue,
urlcolor=blue,
}

\begin{document}
\begin{CJK*}{UTF8}{gbsn} 

\title{Enhanced Atomic Magnetometry with a Pinched Spin State}     
\parbox{\textwidth}

\author{Guo Tingxuan (郭庭轩)}
    \affiliation{Key Laboratory of Nuclear Physics and Ion-Beam Application (MOE), Fudan University, Shanghai 200433, China
   }
    \affiliation{Department of Nuclear Science and Technology, Institute of Modern Physics, Fudan University, Shanghai 200433, China
   }
\author{
    Li Xiangyu (李翔宇)}
    \affiliation{Key Laboratory of Nuclear Physics and Ion-Beam Application (MOE), Fudan University, Shanghai 200433, China
   }
    \affiliation{Department of Nuclear Science and Technology, Institute of Modern Physics, Fudan University, Shanghai 200433, China
   }
\author{Li Yang (李阳)}
    \affiliation{Key Laboratory of Nuclear Physics and Ion-Beam Application (MOE), Fudan University, Shanghai 200433, China
   }
    \affiliation{Department of Nuclear Science and Technology, Institute of Modern Physics, Fudan University, Shanghai 200433, China
   }
\author{Zhao Kaifeng (赵凯锋)}
    \email{Email address: zhaokf@fudan.edu.cn}
    \affiliation{Key Laboratory of Nuclear Physics and Ion-Beam Application (MOE), Fudan University, Shanghai 200433, China
   }
    \affiliation{Department of Nuclear Science and Technology, Institute of Modern Physics, Fudan University, Shanghai 200433, China
   }

\date{\today}

\begin{abstract}
The sensitivity of an atomic spin sensor is fundamentally constrained by the quantum Fisher information (QFI) of the probe state. The pinched state, $|{F,0}\rangle$, has an $(F+1)$-fold larger single-particle QFI for transverse-field sensing than the stretched state, $|{F,F}\rangle$, corresponding to a predicted $\sqrt{F+1}$ reduction in the spin-projection-noise-limited field uncertainty. We show that this advantage can be accessed in an rf sensing scheme by engineering a spectrum symmetric about $m=0$, yielding an ideal $(F+1)$-fold enhancement of magnetic-field response. For $^{87}\mathrm{Rb}$ with $F=2$, we observe up to 2.7-fold response enhancement. Implementing this approach in a compact $0.8~\mathrm{cm}^3$ anti-relaxation-coated multipass cell, we realize a single-beam, dual-axis, zero-field magnetometer, achieving $13$--$23$~$\mathrm{fT}/\sqrt{\mathrm{Hz}}$ over $3$--$100$~Hz at room temperature.
\end{abstract}
\maketitle
\end{CJK*}

Quantum sensing estimates weak physical signals by exploiting the controlled quantum dynamics of a probe system~\cite{degen_rmp_2017,pezze_rmp_2018}. 
In many spin sensors, the signal is encoded as a small rotation of the probe state, and the achievable sensitivity is determined jointly by the quantum fluctuations of the measured observable and its response to the signal. 
For $N$ uncorrelated atoms, spin projection noise (SPN) gives the characteristic $1/\sqrt{N}$ standard-quantum-limit scaling.
A widely explored route to improving sensitivity is spin squeezing, which introduces many-body correlations to reduce the fluctuations of the measured collective-spin component~\cite{pezze_rmp_2018}. 
Spin squeezing has produced substantial metrological gains in cold-atom systems~\cite{muessel_prl_2014,hosten_nature_2016,mao_natphys_2023}. 
In warm atomic vapors, squeezing by quantum nondemolition measurement has been employed in rf magnetometry~\cite{braginsky_science_1980,vasilakis_generation_2015, bao_nature_2020}. The squeezed states are commonly generated around a stretched (spin-coherent) state (SS), $\ket{F,F}$, with the metrological gain benchmarked against the SPN-limited sensitivity of the SS. In practice, however, decoherence and imperfect optical readout strongly limit the attainable gain~\cite{auzinsh_prl_2004,zheng_prl_2023b,duan_natphys_2025}. 

Here we pursue a complementary approach by optimizing the single-particle probe state itself. For a spin-$F>1/2$ system, the stretched state is not the most sensitive separable state, because the additional Zeeman sublevels provide access to states with larger quantum Fisher information (QFI), even without interatomic entanglement. For $N$ uncorrelated atoms in a pure single-particle state, the quantum Cramér-Rao bound (QCRB) for estimating a signal-induced rotation angle, encoded through $U(\theta)=e^{-i\theta G}$, is~\cite{braunstein_prl_1994,pezze_rmp_2018}
\begin{equation}
\label{eq:QCRB}
    {(\delta \theta)}^2 \geq \frac{1}{N I_Q},
\end{equation}
where $I_Q=4{(\delta G)}^2$ is the single-particle QFI and $G$ is the generator of the rotation. For magnetometry, the rotation is generated by the spin component along the measured field. For a magnetic field along $x$, $G=F_x$, and $\theta=\gamma B_x t$, so that $\delta B_x\propto 1/\delta F_x$. 

Within the spin-$F$ manifold, the cat state, $(|F,F\rangle_x+|F,-F\rangle_x)/\sqrt{2}$ maximizes $(\delta F_x)^2$, giving $2F$ times the QFI of the stretched state $|F,F\rangle_z$~\cite{chalopin_natcommun_2018,dietsche_natphys_2019,yang_natphoton_2025}. Such states, however, are difficult to prepare, read out and preserve in warm vapors.

\begin{figure}[htbp] 
    \includegraphics[width=\linewidth]{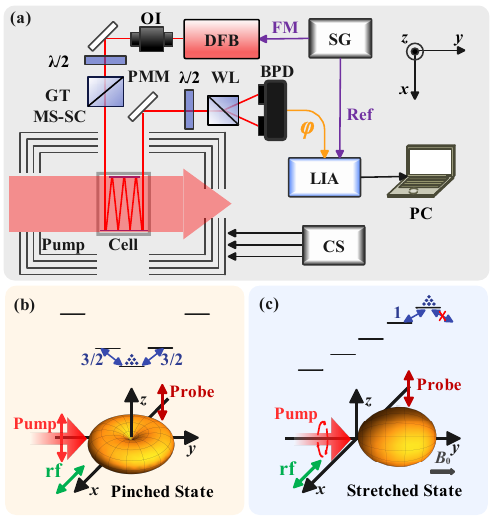}
    \caption{\label{fig:setup} Principle of the rf magnetometer.\ (a): Experimental setup. 
    The laser is frequency modulated (FM) by a signal generator (SG), whose sync signal is fed to the lock-in amplifier (LIA) as reference (Ref) for demodulation.
    OI: optical isolator, $\lambda/2$: half-wave plate, GT: Glan-Thompson prism, PMM: polarization-maintaining mirror, WL: Wollaston prism, BPD: balanced photo-detector, $\varphi$: optical rotation of the probe beam, CS: current source, MS-SC: magnetic shield and static coils, PC: personal computer. 
    (b) and (c): pump, probe and field configurations for the pinched and stretched states, respectively. The orange surface represents each state's angular-momentum probability surface.
    The upper part shows the Zeeman energy-level and transition diagram for each case. The double arrows and the numbers beside them mark the rf-driven transitions and the corresponding coupling strengths, $C_m^2$, obtained from Eq.~\eqref{eq:coupling-strength} for $F=2$.}
\end{figure}

Here we exploit the experimentally accessible high-QFI state $|F,0\rangle_z$, and develop a sensing scheme that converts its large transverse-spin variance into an enhanced magnetic-field response. We refer to this state as the \emph{pinched state} (PS) because its angular-momentum probability surface resembles a sphere pinched at the poles (see Fig.\ref{fig:setup} (b))~\cite{rochester_ajp_2001b}. For integer $F$, $\ket{F,0}_z$ has the largest transverse-spin variance among the Zeeman eigenstates,
\begin{equation}
\label{eq:VariancePinched}
    {(\delta F_x)}_\mathrm{PS}=\sqrt{\frac{F(F+1)}{2}},
\end{equation}
which is a factor of $\sqrt{F+1}$ larger than that of the stretched state,
\begin{equation}
\label{eq:VarianceStretched}
    {(\delta F_x)}_\mathrm{SS}=\sqrt{\frac{F}{2}}.     
\end{equation}
Thus, at the spin-projection-noise limit, the pinched state permits a $\sqrt{F+1}$-fold lower field uncertainty than the stretched state, without spin squeezing.

The QCRB does not by itself specify a measurement that attains this advantage. For a chosen readout observable $O$, error propagation gives
\begin{equation}
\label{eq:SensitivityReadout}
\delta\theta = \frac{\delta O}{\left|\partial\langle O\rangle/\partial\theta\right|},   
\end{equation}
where $\delta O$ is the uncertainty of the readout observable and $\partial\langle O\rangle/\partial\theta$ is its response slope. For the conventional stretched-state magnetometer, $F_x$ is the readout observable. Because the pinched state's $F_x$ noise is larger by $\sqrt{F+1}$, using the same $F_x$ readout requires an $F+1$-fold response enhancement to realize its QFI advantage. We achieve this by engineering a spectrum symmetric about $m=0$.

We consider an rf magnetometer based on a spin system with gyromagnetic ratio $\gamma$  driven by $\mathbf{B}_{\rm rf}=B_{\rm rf}\cos(\omega t)\,\hat{\mathbf{x}}$. The interaction Hamiltonian is 
\begin{equation}
    \label{eq:rfHamiltonian}
    H_{\rm rf}=\hbar\Omega_{\rm rf}F_x\cos(\omega t),
\end{equation}
where $\Omega_{\rm rf}=\gamma B_{\rm rf}$. A static Hamiltonian $H_0$ defines the Zeeman eigenstates and energies $E_m$, with adjacent-state transition frequencies $\omega_m=(E_m-E_{m-1})/\hbar$. When $|\omega_{m}|$ is tuned to match $\omega$, the rf field resonantly drives coherence between adjacent Zeeman states with coupling strength $C^2_{m}$, where
\begin{equation}
C_{m}\equiv \bra{m}F_x\ket{m-1}=\frac{\sqrt{F(F+1)-m(m-1)}}{2}.
\label{eq:coupling-strength}
\end{equation}
This produces an oscillating transverse spin orientation $\langle F_x\rangle$ that can be read out optically. 

We describe the spin dynamics with the density matrix master equation:
\begin{equation}
    \label{eq:Master}
    \frac{d\rho}{dt} =-\frac{i}{\hbar}\left[H_{0}+H_{\rm rf},\rho
    \right]-\Gamma (\rho-\rho^{0}),
\end{equation}
where $\Gamma$ is a phenomenological relaxation rate, and $\rho^{0}$ is the rf-free steady-state density matrix. 

A conventional rf magnetometer uses a magnetic field along $z$ to produce a Hamiltonian linear in $F_z$,
\begin{equation}
    \label{eq:LinearHamiltonian}
    H_{\rm 0}=\hbar \Omega_{\rm L}F_z,
\end{equation}
giving $E_m=m\hbar\Omega_{\rm L}$ and $\omega_m=\Omega_{\rm L}$. For the stretched state $\ket{F,F}$, $\ket{F,F}\leftrightarrow\ket{F,F-1}$ transition contributes, with $C^2_{F}=F/2$, and only the co-rotating circular component of the rf field is resonant~(see Fig.~\ref{fig:setup}(c)). In the weak-drive regime, $\Omega_\mathrm{rf}\ll \Gamma$, the steady-state oscillation amplitude of $\langle F_x \rangle$ is
\begin{equation}
{|\ev{ F_x}|}=
\frac{F\Omega_\mathrm{L}\Omega_\mathrm{rf} }{\sqrt{\Gamma^4+{(\omega^2-\Omega_\mathrm{L}^2)}^2+2\Gamma^2(\omega^2+\Omega_\mathrm{L}^2)}}.
\label{eq:ResponseStretched}
\end{equation}

For the pinched state, we can engineer a spectrum that is symmetric about the populated $m=0$ state with a Hamiltonian quadratic in $F_z$,
\begin{equation}
    H_{0}=\hbar\Omega_{\rm TS}F_z^2,
\end{equation}
giving $\omega_m=(2m-1)\Omega_\mathrm{TS}$. 
Such $H_0$ can be realized by a tensor light shift~\cite{mathur_pr_1968}, microwave dressing~\cite{gerbier_pra_2006,luo_science_2017b}, or an oscillating magnetic field~\cite{dezhangfard_prap_2025b}.
In this case, the two circular components of the rf field can simultaneously drive the $\ket{F,0}\leftrightarrow\ket{F,\pm1}$ transitions, which also have the largest coupling strengths in the manifold, with $C^2_{0}=C^2_{1}=F(F+1)/4$ (see Fig.~\ref{fig:setup}(b)). For an ideal pinched state in the weak-drive regime, the steady-state oscillation amplitude of $\langle F_x \rangle$ is
\begin{equation}\label{eq:ResponsePinched}
{|\ev{ F_x}|}=
\frac{F(F+1)\Omega_\mathrm{TS}\Omega_\mathrm{rf}}{\sqrt{\Gamma^4+{(\omega^2-\Omega_\mathrm{TS}^2)}^2+2\Gamma^2(\omega^2+\Omega_\mathrm{TS}^2)}}.
\end{equation}

Therefore, for equal relaxation rates and matched resonance frequencies, $\Omega_{\rm TS}=\Omega_{\rm L}$, the pinched state enhances the field response by a factor of $F+1$ relative to the stretched state, as detailed in the Supplemental Material~\cite{SM}. This response enhancement of the pinched state outweighs its $\sqrt{F+1}$-fold increase in projection noise, yielding a predicted $\sqrt{F+1}$ reduction in SPN-limited sensitivity relative to the stretched state. When state-independent readout noise dominates, the larger response alone can instead yield a sensitivity improvement approaching $F+1$.


The coherent conversion of pinched-state alignment into orientation is a resonant form of alignment-to-orientation conversion (AOC)~\cite{Hilborn_pra_1994,alexandrov_josab_2005}. Unlike conventional AOC magnetometry~\cite{budker_prl_2000,budker_rmp_2002,kimball_jap_2009,bevington_pra_2020}, we prepare the alignment in the pinched state and operate in the resolved-splitting regime, $\Omega_{\rm TS}>\Gamma$, where the rf field resonantly drives the tensor-split transitions.

Next we test the predicted response enhancement using the setup in Fig.~\ref{fig:setup}. The sensor head is a planar multipass, alkene-coated $^{87}\mathrm{Rb}$ vapor cell~\cite{li_arxiv_2026}, with internal dimensions of $8\times8\times12\ \mathrm{mm^3}$ (cross-section$\times$length).
The cell is positioned at the center of a four-layer cylindrical $\mu$-metal magnetic shield. 
Bias and calibration fields are produced by three orthogonal coils driven by a low-noise three-channel current source~\cite{mrozowski_rsi_2023}. 
A $z$-polarized 5 $\mu\mathrm{W}$ probe propagates along $x$, traversing the cell 12 times with $60\%$ transmission. We label the $F=I+1/2$ and $F=I-1/2$ ground-state hyperfine levels as $F_a$ and $F_b$ respectively, and the corresponding D1 excited-state levels as $F_a'$ and $F_b'$.
The laser is frequency-modulated at $\omega_{\mathrm{FM}}/2\pi=20$ kHz over detunings 5--11 GHz red of the $F_a \rightarrow F_b'$ transition. The Faraday rotation  $\varphi \propto \ev{F_x}$ is detected with a polarimeter and a lock-in amplifier demodulates at $\omega_{\mathrm{FM}}$.

A pump beam, consisting of a direct-pump component near the D1 $F_a \rightarrow F_a'$ transition and an indirect-pump component at the $F_b \rightarrow F_a'$ resonance, propagates along $y$ and prepares the atomic ensemble in the target state within the $F_a$ manifold. 
To prepare the pinched state, the combined pump beam is polarized along $z$. The direct component is blue-detuned by several hundred MHz, producing the tensor light shift. 
The stretched state is produced using resonant, circularly polarized direct and indirect pumps in the presence of a bias field $\mathbf{B}_0$ along $y$. Constraints imposed by the shield geometry require the stretched state to be oriented along $y$ rather than $z$. However, since the $y$- and $z$-oriented configurations related by a rotation about the $x$ axis, Eq.\eqref{eq:ResponseStretched} remains valid for the $F_x$ response. We adjust the direct-pump power to prepare the desired state while matching the relaxation rates of the pinched- and stretched-state resonances.
\begin{figure}[htbp] 
    \includegraphics[width=\linewidth]{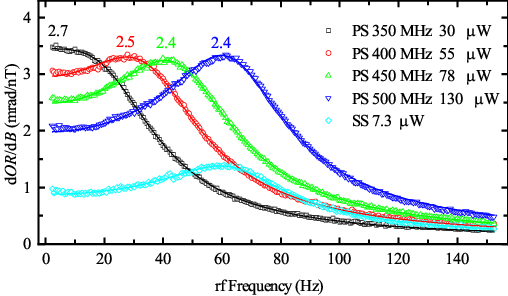}
    \caption{\label{CSSvsPS} Comparison of the rf-field response between the pinched state (PS) and the stretched state (SS) by sweeping the same rf field ($48.5~\mathrm{pT_{pk}}$) along $x$ from 0.5 to 152 Hz. The same 5 $\mathrm{\mu W}$ probe beam, far-detuned on the red wing of the $F_a \rightarrow F_b'$ transition, detects $\langle F_x \rangle$ via Faraday rotation. For the PS scheme, different combinations of power and detuning of the direct pump component are used to shift the resonance frequency while keeping resonance linewidth approximately constant. For the SS scheme, the power of the direct pump is adjusted to match the linewidth of the PS. For both states the power of the on-resonance indirect pump is 80 $\mathrm{\mu W}$.}
\end{figure}

The frequency-dependent field responses of the two states are shown in Fig.~\ref{CSSvsPS}.
The PS response exceeds the SS response by  $2.7\pm 0.1$ at low resonance frequencies, decreasing to about 2.4 at higher frequency. The deviation from the ideal factor of 3 and its frequency dependence arises mainly because the direct-pump beam provides both optical pumping and tensor splitting: increasing its detuning and power raises $\Omega_{\mathrm{TS}}$ but also increases excitation of the undesired $F_a \rightarrow F_b'$ transition, degrading pumping into $\ket{F,0}$. Therefore, separating state preparation from generation of the quadratic splitting through, for example, microwave dressing or an additional light-shift beam should bring the response closer to the ideal enhancement.

The linewidth is presently dominated by optical pumping and could be substantially reduced by temporally separating the pumping and probing stages. However, compared with the stretched state, the pinched state is more susceptible to spin-exchange relaxation. Room-temperature operation provides a favorable compromise between sensitivity and bandwidth.

To demonstrate the practical utility of the pinched-state protocol, we implement a single-beam, dual-axis, broadband zero-field rf magnetometer. 
We use the same setup as that shown in Fig.~\ref{fig:setup} except that the separate pump beam is removed.
The probe frequency is square-wave modulated at $\omega_{\mathrm{FM}}/2\pi=20\ \mathrm{kHz}$, alternating between the blue wing of the $F_a \rightarrow F_a'$ transition and the center of the $F_b \rightarrow F_a'$ transition. This modulated beam simultaneously prepares the pinched state, generates the tensor light shift, and probes the rf-field-induced spin orientation through the optical rotation $\varphi$ using phase-sensitive demodulation at $\omega_{\mathrm{FM}}$.

\begin{figure}[htbp]
    \includegraphics[width=1.0\linewidth]{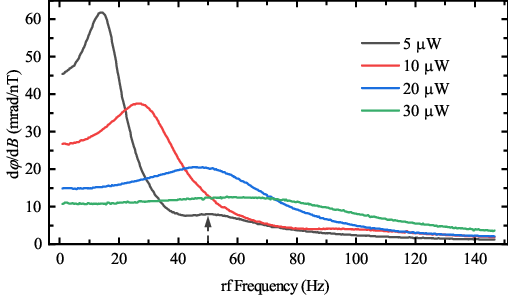}
    \caption{\label{response2} Frequency-dependent field response of the single beam PS magnetometer at a fixed center detuning for different probe powers at 23 ${^{\circ}\mathrm{C}}$. The arrow indicates the location of the $3\Omega_{\mathrm{TS}}$ resonance associated with the $\ket{2,\pm 1} \leftrightarrow \ket{2,\pm 2}$ transitions at insufficient pumping.}
\end{figure}

Fig.~\ref{response2} shows the field response at fixed probe detuning for several probe powers. 
Increasing probe power shifts $\Omega_{\mathrm{TS}}$ and broadens the resonance, consistent with tensor-light-shift generation and optical-pumping-dominated relaxation. At lower powers, a weak  $3\Omega_{\mathrm{TS}}$ resonance from the $|2,\pm1\rangle \leftrightarrow |2,\pm2\rangle$ transitions reveals incomplete optical pumping into $|2,0\rangle$.

A field $B_y$ along the $y$-axis would similarly generate a spin orientation along $y$. To detect $B_y$, a longitudinal field $B_z^{\mathrm{mod}}$ modulated at $\omega_z/2\pi =2~\mathrm{kHz}$ converts the induced $y$-orientation into $x$-orientation detected by the probe; demodulation at $\omega_z$ yields the $y$ channel. The modulation depth of $B_z^{\mathrm{mod}}$ is set to 0.8 rad to balance the field response of the two channels.

\begin{figure}[htbp]
    \includegraphics[width=1.0\linewidth]{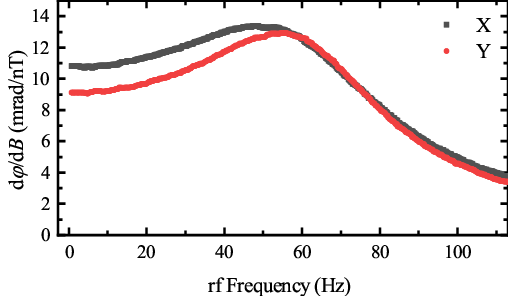}
    \caption{\label{response3} Frequency-dependent field response of the dual-axis configuration at 23 ${^{\circ}\mathrm{C}}$. Probe power is 22 $\mu\mathrm{W}$, and the lower-frequency endpoint of the laser-frequency modulation is at least 400~MHz blue-detuned from the $F_a \rightarrow F_a'$ transition.}
\end{figure}

Guided by Fig.~\ref{response2}, we choose a probe power of 22 $\mu\mathrm{W}$ and set the lower-frequency endpoint of the laser-frequency modulation at least \(400~\mathrm{MHz}\) blue of the $F_a \rightarrow F_a'$ transition, balancing sensitivity and bandwidth. The resulting frequency-dependent field responses of the two channels are shown in Fig.~\ref{response3}.
The difference in $\Omega_{\mathrm{TS}}$ and $\Gamma$ between the two channels may arise from the considerable linear dichroism in the $x$-channel.

\begin{figure}[htbp]
    \includegraphics[width=1.0\linewidth]{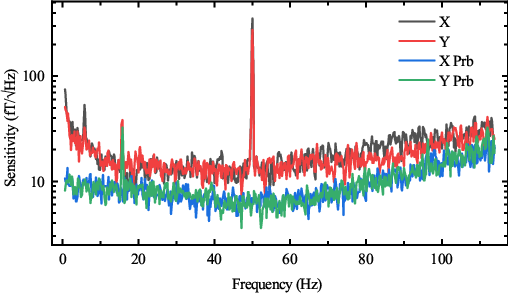}
    \caption{\label{sensitivity2} Noise spectrum of the magnetometer. The black and red lines represent the noise background for the $x$ and $y$ channels, respectively, in the absence of the calibration field. The blue and green lines respectively represent light noise for the $x$ and $y$ channels, which are taken by blue-detuning the frequency-modulation center of the laser 35 GHz from the D1 center while maintaining the same optical power. All data are taken at 23 ${^{\circ}\mathrm{C}}$ and calibrated by the response curve measured in Fig.~\ref{response3}.}
\end{figure}

To measure the magnetometer's sensitivity, calibration fields at 60 Hz with known amplitudes are applied separately along the $x$ (34.3 $\mathrm{pT_{rms}}$) and $y$ (25.8 $\mathrm{pT_{rms}}$) axes. 
The noise floor measured in the absence of the calibration field is shown in Fig.~\ref{sensitivity2}. Excluding narrow technical spectral lines, the magnetometer achieves baseline sensitivities of $13$--$23~\mathrm{fT/\sqrt{Hz}}$ across $3$--$100~\mathrm{Hz}$ on both measurement axes.
The on-resonance photon shot noise, $\delta B_{\mathrm{PSN}}$, is estimated to be $6~\mathrm{fT/\sqrt{Hz}}$ using the probe power detected by the photo-detector and the response curve in Fig.~\ref{response3}. The corresponding spin-projection-noise limit, $\delta B_{\mathrm{SPN}}$, is estimated to be $4~\mathrm{fT/\sqrt{Hz}}$~\cite{SM,savukov_prl_2005,tang_pra_2020}, giving a combined PSN and SPN noise floor of approximately $7~\mathrm{fT/\sqrt{Hz}}$ at resonance.
The remaining excess noise is consistent with magnetic noise from the $\mu$-metal shield and probe-induced noise. Because the present magnetometer operates above the SPN limit, the predicted $\sqrt{F+1}$ improvement in SPN-limited sensitivity is not directly resolved in this experiment.
The achieved performance is comparable to the state-of-the-art commercial SERF magnetometers~\cite{quspin_2026}, despite operation at room temperature  with more than two orders of magnitude fewer atoms. 
Our magnetometer further combines a compact sensing volume with broadband dual-axis vector detection, distinguishing it from other room-temperature zero-field magnetometers with comparable sensitivity~\cite{kimball_jap_2009,qu_nc_2020,Fourcault_oe_2021}.

In conclusion, the pinched state, $|F,0\rangle$, combined with a quadratic spin spectrum, provides an ideal $F+1$-fold enhancement of magnetic-field response and a predicted $\sqrt{F+1}$ improvement in SPN-limited sensitivity relative to the stretched state. For $^{87}\mathrm{Rb}$ with $F=2$, we observe up to 2.7-fold response enhancement and realize a single-beam, dual-axis zero-field magnetometer, achieving 13--23~$\mathrm{fT/\sqrt{Hz}}$ over 3--100 Hz in a $0.8~\mathrm{cm^3}$ cell at room temperature. These results demonstrate a practical route for converting high single-particle QFI into enhanced measurable response, with the potential for improved SPN-limited sensitivity without spin squeezing.

\FloatBarrier     

\begin{acknowledgments}
We acknowledge support from the National Natural Science Foundation of China (Major Scientific Research Instrument Development Project No. 12027806). 
\end{acknowledgments}

\section*{Data availability}
There are no publicly available research data or software supporting this manuscript. Requests for further information or data should be sent to the authors.

\bibliography{reference.bib}
\end{document}


\begin{CJK*}{UTF8}{gbsn} 
\title{Supplemental Material: Enhanced Atomic Magnetometry with a Pinched Spin State}

\author{Guo Tingxuan (郭庭轩)}
    \affiliation{Key Laboratory of Nuclear Physics and Ion-Beam Application (MOE), Fudan University, Shanghai 200433, China
   }
    \affiliation{Department of Nuclear Science and Technology, Institute of Modern Physics, Fudan University, Shanghai 200433, China
   }
\author{
    Li Xiangyu (李翔宇)}
    \affiliation{Key Laboratory of Nuclear Physics and Ion-Beam Application (MOE), Fudan University, Shanghai 200433, China
   }
    \affiliation{Department of Nuclear Science and Technology, Institute of Modern Physics, Fudan University, Shanghai 200433, China
   }
\author{Li Yang (李阳)}
    \affiliation{Key Laboratory of Nuclear Physics and Ion-Beam Application (MOE), Fudan University, Shanghai 200433, China
   }
    \affiliation{Department of Nuclear Science and Technology, Institute of Modern Physics, Fudan University, Shanghai 200433, China
   }
\author{Zhao Kaifeng (赵凯锋)}
    \email{Email address: zhaokf@fudan.edu.cn}
    \affiliation{Key Laboratory of Nuclear Physics and Ion-Beam Application (MOE), Fudan University, Shanghai 200433, China
   }
    \affiliation{Department of Nuclear Science and Technology, Institute of Modern Physics, Fudan University, Shanghai 200433, China
   }

\date{\today}
\maketitle
\end{CJK*}
\section{A Spin System under Tensor Light Shift, Zeeman Shift, and a Weak Linearly Polarized Transverse RF Drive}
\subsection{Hamiltonian and Master Equation}
A spin-$F$ system with gyromagnetic ratio $\gamma$ is subject to a static field $B_0\hat{\mathbf z}$, a tensor light shift $\hbar\Omega_{\rm TS}F_z^2$, and an rf field along $x$, with $\Omega_{\rm L}=\gamma B_0$ and
$\Omega_{\rm rf}=\gamma B_{\rm rf}$. The Hamiltonian is 
\begin{equation}
H(t)=\hbar \Omega_\mathrm{L} F_z +\hbar\Omega_\mathrm{TS} F_z^2+\hbar\Omega_{\rm rf}\cos(\omega t)F_x,
\end{equation}

Assuming a uniform relaxation rate $\Gamma$, the density matrix obeys the master equation
\begin{equation}
\frac{d\rho}{dt}=-\frac{i}{\hbar}[H(t),\rho]-\Gamma(\rho-\rho^{(\mathrm{0})}).
\label{eq:master}
\end{equation}
Here $\rho^{(0)}$ is the rf-free steady state. In the Zeeman basis $|m\rangle$, the adjacent matrix elements of $F_x$ are
\begin{equation}\label{eq:Cm}
    C_m\equiv\mel{m}{F_x}{m-1}=\frac{\sqrt{F(F+1)-m(m-1)}}{2},
\end{equation}
with 
\begin{equation}\label{eq:Cm2}
    C_m^2=C_{1-m}^2.
\end{equation}
The transition frequencies are
\begin{equation}
\omega_m \equiv \frac{E_m-E_{m-1}}{\hbar} = \Omega_\mathrm{L} + (2m-1)\Omega_\mathrm{TS},
\end{equation}
where $E_m$ is the eigenenergy corresponding to the eigenstate $\ket{m}$ of the static Hamiltonian. 

We define the detuning $\Delta_m$ and sum frequency $\Sigma_m$ as:
\begin{equation}
\Delta_m \equiv \omega-\omega_m = \omega-\Omega_\mathrm{L}- (2m-1)\Omega_\mathrm{TS},
\end{equation}
\begin{equation}
\Sigma_m \equiv \omega+\omega_m = \omega+\Omega_\mathrm{L} + (2m-1)\Omega_\mathrm{TS}.
\end{equation}
We also define the adjacent population differences as
\begin{equation}
D_m\equiv \rho_{mm}-\rho_{m-1,m-1}.
\end{equation}

\subsection{Weak-Drive Approximation}
To first order in the weak-drive, $\Omega_{\rm rf}\ll \Gamma$, 
\begin{equation}
\dot\rho_{mm}=-\Gamma(\rho_{mm}-\rho_{mm}^{(\mathrm{0})}),
\label{eq:rho-mm}
\end{equation}
\begin{equation}
\begin{aligned}
\dot\rho_{m,m-1}=-(\Gamma+i\omega_m)\rho_{m,m-1}+i\Omega_\mathrm{rf}C_m D_m\cos(\omega t).
\label{eq:coh-eq}
\end{aligned}
\end{equation}
For the steady-state populations, we have 
\begin{equation*}
\rho_{mm}\approx \rho_{mm}^{(\mathrm{0})},
\qquad
D_m\approx D_m^{(\mathrm{0})}.
\end{equation*}
To obtain the steady-state solution to Eq.~\eqref{eq:coh-eq}, we write $\cos(\omega t)=\frac{1}{2}\left(e^{i\omega t}+e^{-i\omega t}\right)$ and use the ansatz
\begin{equation}
\rho_{m,m-1}(t)=a_m e^{-i\omega t}+b_m e^{i\omega t},
\end{equation}
Substitution into Eq.~\eqref{eq:coh-eq} gives
\begin{equation}
a_m=\frac{\Omega_{\rm rf} C_m D_m}{2}\frac{i}{\Gamma-i\Delta_m},
\qquad
b_m=\frac{\Omega_{\rm rf} C_m D_m}{2}\frac{i}{\Gamma+i\Sigma_m}.
\end{equation}
Thus,
\begin{equation}\label{eq:Fx-general}
    \begin{aligned}
        \langle F_x(t)\rangle &=\sum_{m=1-F}^{F}2C_m\Re\rho_{m,m-1}(t) ={\Omega_{\rm rf}}\sum_{m=1-F}^{F} C_m^2 D_m (A_m+B_m),    
    \end{aligned}
\end{equation}
where $A_m$ and $B_m$ are defined as
\begin{equation} 
    \begin{gathered}
A_m\equiv \frac{-\Delta_m\cos\omega t+\Gamma\sin\omega t}{\Gamma^2+\Delta_m^2},
\qquad
B_m\equiv \frac{\Sigma_m\cos\omega t-\Gamma\sin\omega t}{\Gamma^2+\Sigma_m^2}.
\end{gathered}
\end{equation} 
Similarly, 
\begin{equation}\label{eq:Fy-general}
    \begin{aligned}
\langle F_y(t)\rangle&=-\sum_{m=1-F}^{F}2C_m\Im\rho_{m,m-1}(t)={\Omega_{\rm rf}}\sum_{m=1-F}^{F} C_m^2 D_m (A'_m+B'_m),
    \end{aligned}
\end{equation}
where $A'_m$ and $B'_m$ are defined as
\begin{equation} 
    \begin{gathered}
A'_m \equiv -\frac{\Delta_m\sin\omega t+\Gamma\cos\omega t}{\Gamma^2+\Delta_m^2} 
\qquad
B'_m\equiv -\frac{\Sigma_m\sin\omega t+\Gamma\cos\omega t}{\Gamma^2+\Sigma_m^2}.
\end{gathered}
\end{equation}

\subsection{Linearly Polarized Optical Pumping under Zero Bias Field}
We first consider the case of linearly polarized optical pumping under zero bias field. In this case,  $\Omega_\mathrm{L}=0$, thus
\begin{equation}\label{eq:Deltam}
    \Delta_m=\Sigma_{1-m}=\omega-(2m-1)\Omega_\mathrm{TS}.
\end{equation} 
The rf-free steady-state density matrix is symmetric with respect to $m\rightarrow -m$, i.e., $\rho_{mm}=\rho_{-m,-m}$, which implies that
\begin{equation}\label{eq:Dm}
    D_m= -D_{1-m}.
\end{equation}

\subsubsection{Vector Response}
Applying Eqs.~\eqref{eq:Deltam},~\eqref{eq:Dm}, and~\eqref{eq:Cm2} to Eq.~\eqref{eq:Fy-general}, one finds 
\begin{equation}
    C_m^2 D_m A'_m = - C_{1-m}^2 D_{1-m}B'_{1-m}.    
\end{equation}
As a result, in Eq.~\eqref{eq:Fy-general}, the terms from the $(m, 1-m)$ pair cancel pairwise, yielding
\begin{equation}
\langle F_y(t)\rangle=0.
\end{equation}

Applying Eqs.~\eqref{eq:Deltam},~\eqref{eq:Dm}, and~\eqref{eq:Cm2} to Eq.~\eqref{eq:Fx-general}, one finds
\begin{align}
    C_m^2 D_m A_m = C_{1-m}^2 D_{1-m}B_{1-m}.
\end{align}
As a result, $\langle F_x(t)\rangle$ is simplified to
\begin{equation}\label{eq:FxAlignment}
\langle F_x(t)\rangle=2{\Omega_{\rm rf}}\sum_{m=1-F}^{F} C_m^2 D_m A_m
=2{\Omega_{\rm rf}}
\sum_{m=1-F}^{F} C_m^2 D_m
\frac{-\Delta_m\cos\omega t+\Gamma\sin\omega t}{\Gamma^2+\Delta_m^2}.
\end{equation}
The corresponding amplitude ${|\langle F_x\rangle|}$ is 
\begin{equation}\label{eq:AmpxAlignment}
  {|\langle F_x\rangle|}=2{\Omega_{\rm rf}}
\sqrt{{\left(\sum_{m=1-F}^{F} C_m^2 D_m \frac{\Delta_m}{\Gamma^2+\Delta_m^2}\right)}^2
+{\left(\sum_{m=1-F}^{F} C_m^2 D_m \frac {\Gamma}{\Gamma^2+\Delta_m^2}\right)}^2}.
\end{equation}

By rotational symmetry about $z$, a $y$-directed rf field analogously produces only a $y$-oriented response.
Thus the induced orientation is parallel to the transverse rf field.

\subsubsection{The Pinched State}
For the pinched state (PS), with all population in $m=0$,
$D_0=-D_1=1$ and all other $D_m=0$. Equation~\eqref{eq:Fx-general} becomes 
\begin{equation}
\begin{aligned}
\langle F_x(t)\rangle
=\frac{\Omega_{\rm rf}}{2}F(F+1)
\left[
\frac{(\omega-\Omega_\mathrm{TS})\cos\omega t-\Gamma\sin\omega t}{\Gamma^2+{(\omega-\Omega_\mathrm{TS})}^2}
-
\frac{(\omega+\Omega_\mathrm{TS})\cos\omega t-\Gamma\sin\omega t}{\Gamma^2+{(\omega+\Omega_\mathrm{TS})}^2}
\right].
\label{eq:FxPinched}
\end{aligned}
\end{equation}
The corresponding amplitude [Fig.~\ref{fig:CompareSS&PS}\subref{fig:sub1}] is
\begin{equation}\label{eq:AmpxPinched}
  {|\langle F_x\rangle|}=\frac{F(F+1)\Omega_\mathrm{TS} }{\sqrt{\Gamma^4+{(\omega^2-\Omega_\mathrm{TS}^2)}^2+2\Gamma^2(\omega^2+\Omega_\mathrm{TS}^2)}}\Omega_{\rm rf}.
\end{equation}

\subsection{Stretched State under Zero Tensor Light Shift}
\begin{figure}[htbp]
    \centering
    \captionsetup[subfigure]{skip=-9pt}
    \begin{subfigure}[b]{0.49\textwidth}
        \centering
        \includegraphics[width=\linewidth]{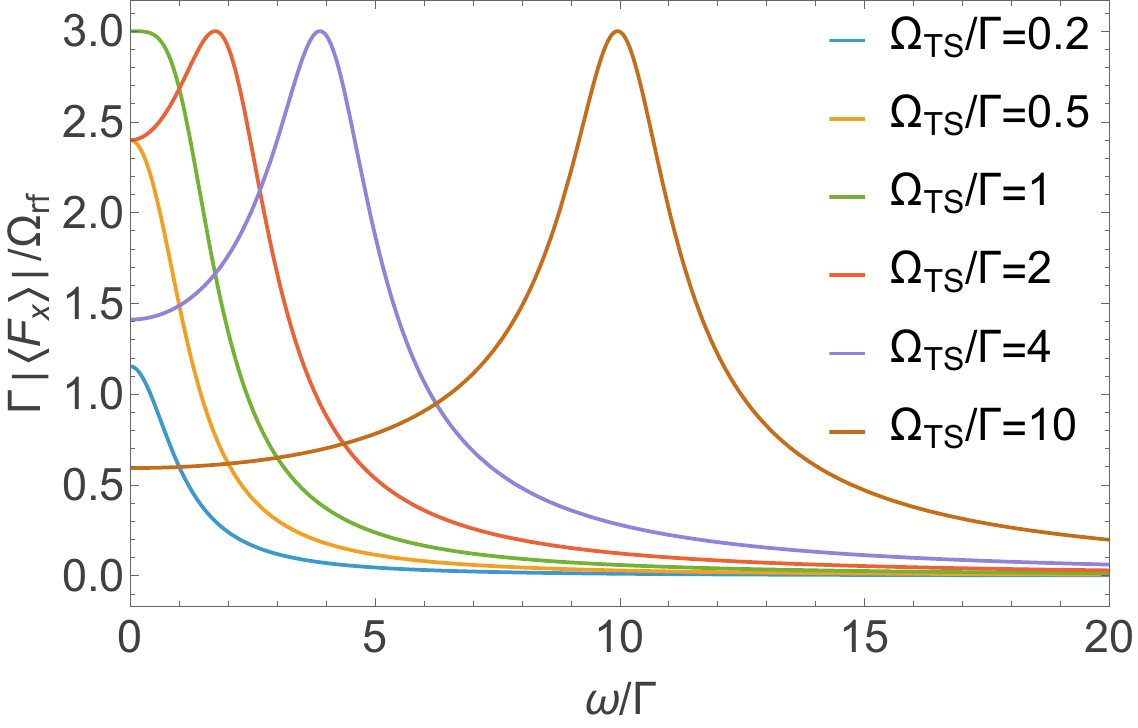}
        \caption{}
        \label{fig:sub1}
    \end{subfigure}
    \hfill
    \begin{subfigure}[b]{0.49\textwidth}
        \centering
        \includegraphics[width=\linewidth]{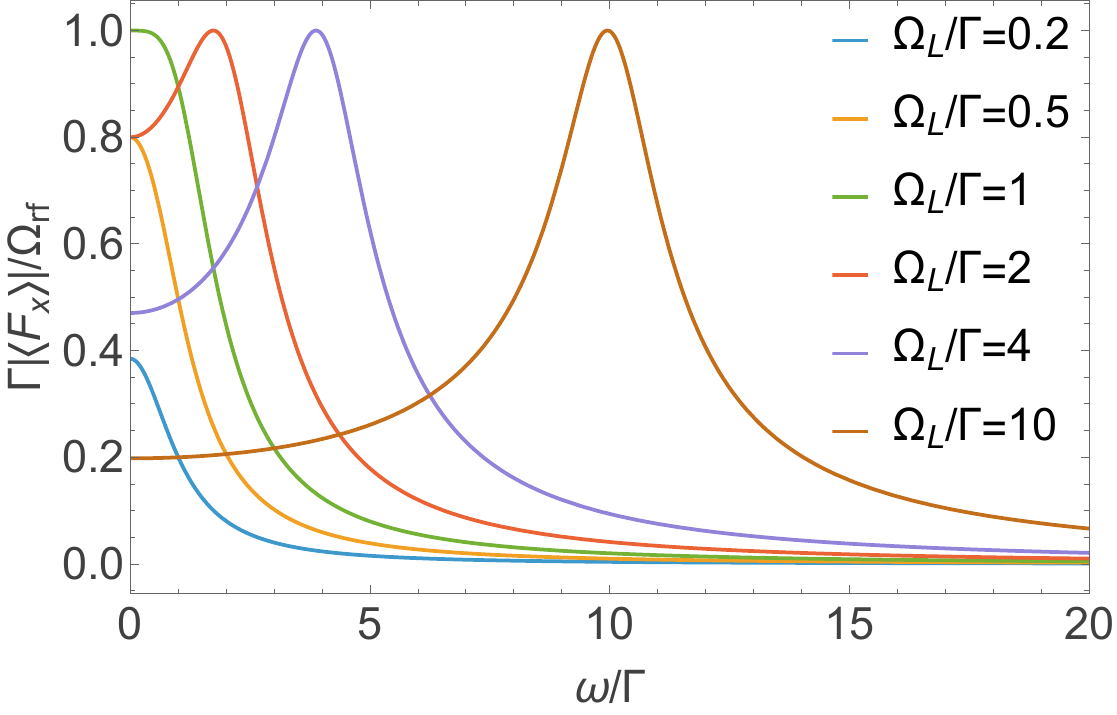}
        \caption{}
        \label{fig:sub2}
    \end{subfigure}
    \captionsetup{format=plain,font=small,justification=raggedright}
    \caption{Normalized rf response for $F=2$.
(a) PS for several $\Omega_{\rm TS}/\Gamma$.
(b) SS for several $\Omega_{\rm L}/\Gamma$.
Eqs.~\eqref{eq:Ampx0Pinched} and ~\eqref{eq:Ampx0Stretched} apply for
$\Omega_{\rm TS}\geq\Gamma$ and $\Omega_{\rm L}\geq\Gamma$, respectively.} 
    \label{fig:CompareSS&PS}
\end{figure}

For the stretched state (SS) with $\Omega_{\rm TS}=0$,
$D_F=1$ and all other $D_m=0$. Equation~\eqref{eq:Fx-general} becomes
\begin{equation}\label{eq:FxStretched}
    \begin{aligned}
        \langle F_x(t)\rangle &=\frac{\Omega_{\rm rf}}{2}F\left[
\frac{-(\omega-\Omega_\mathrm{L})\cos\omega t+\Gamma\sin\omega t}{\Gamma^2+{(\omega-\Omega_\mathrm{L})}^2}
+
\frac{(\omega+\Omega_\mathrm{L})\cos\omega t-\Gamma\sin\omega t}{\Gamma^2+{(\omega+\Omega_\mathrm{L})}^2}
\right].    
    \end{aligned}
\end{equation}
The corresponding amplitude [Fig.~\ref{fig:CompareSS&PS}\subref{fig:sub2}] is
\begin{equation}\label{eq:AmpxStretched}
  {|\langle F_x\rangle|}=\frac{F\Omega_\mathrm{L} }{\sqrt{\Gamma^4+{(\omega^2-\Omega_\mathrm{L}^2)}^2+2\Gamma^2(\omega^2+\Omega_\mathrm{L}^2)}}\Omega_{\rm rf}.
\end{equation}

For equal $\Gamma$ and matched splittings,
$\Omega_{\rm TS}=\Omega_{\rm L}$, Eqs.~\eqref{eq:AmpxPinched} and \eqref{eq:AmpxStretched} give a response enhancement of $F+1$. This follows from $C_0^2=F(F+1)/4$ versus $C_F^2=F/2$, together with the contribution
of both rotating components for the PS.

\subsection{Resonant Field Response and Spin-Projection Noise}
For the PS with $\Omega_{\rm TS}\geq\Gamma$, Eq.~\eqref{eq:AmpxPinched} is maximized at
$\omega=\sqrt{\Omega_{\rm TS}^2-\Gamma^2}$, giving
\begin{equation}\label{eq:Ampx0Pinched}
{|\langle F_x\rangle|}_0 = \frac{F(F+1)\gamma}{2\Gamma}B_{\rm rf}.
\end{equation}
For the SS with $\Omega_{\rm L}\geq\Gamma$, Eq.~\eqref{eq:AmpxStretched} similarly gives the resonant amplitude at $\omega=\sqrt{\Omega_\mathrm{L}^2-\Gamma^2}$, 
\begin{equation}\label{eq:Ampx0Stretched}
{|\langle F_x\rangle|}_0=\frac{F\gamma}{2\Gamma} B_{\rm rf}.
\end{equation}

Since $\langle F_x \rangle=0$ for both states, the transverse spin variance is
\begin{equation}
(\Delta F_x)^2=\langle F_{x}^2\rangle = \frac{1}{2}\left[F(F+1)-\sum_m m^2\rho_{mm} \right] .
\end{equation}
For $N$ atoms whose transverse-spin correlations decay exponentially
at rate $\Gamma$, the one-sided resonant SPN-equivalent field-noise
amplitude spectral density is~\cite{savukov_prl_2005}
\begin{equation}
    \delta B_{\mathrm{SPN}} = \frac{\partial B_{\rm rf}}{\partial|\langle F_{x}\rangle|}_0 \frac{2\sqrt{\langle F_{x}^2\rangle}}{\sqrt{\Gamma N}}.
\end{equation}  
For the stretched state, $\langle F_x^2\rangle= F/2$. Using Eq.~\eqref{eq:Ampx0Stretched}, we obtain
\begin{equation}
    \delta B_{\mathrm{SPN}} = \frac{2}{\gamma}\sqrt{\frac{2\Gamma}{F N}}.    
\end{equation}  
For the pinched state, $ \langle F_x^2\rangle =F(F+1)/2$. Using Eq.~\eqref{eq:Ampx0Pinched}, we obtain
\begin{equation}
    \delta B_{\mathrm{SPN}} =  \frac{2}{\gamma}\sqrt{\frac{2\Gamma}{F(F+1)N}}. 
\label{eq:SPNPinched}
\end{equation}
Thus, the pinched-state rf magnetometer improves the SPN-limited sensitivity by a factor of $\sqrt{F+1}$ relative to the stretched-state rf magnetometer.

The comparison assumes similar transverse relaxation rates for the PS
and SS. At room temperature,
$\Gamma_{\rm SE}/2\pi\simeq1.3~\mathrm{Hz}$ is much smaller than the
measured $\Gamma/2\pi\simeq20$--$27~\mathrm{Hz}$, so spin exchange
does not significantly affect the PS--SS comparison.

Because the multipass geometry samples a large fraction of the cell and
the spin-relaxation time greatly exceeds the beam transit time, the
narrow Ramsey component dominates the resonant spin noise; thus
$N$ may be approximated by the total atom number~\cite{tang_pra_2020}.
For the single-beam operating point used for the sensitivity measurement in Fig.~5 of the main text, the atom number is estimated from the Rb vapor density at \(23^\circ\mathrm C\) and the cell volume to be $N=8\times 10^9$. The fitted relaxation rate is $\Gamma/2\pi = 27~\mathrm{Hz}$, and Eq.~\eqref{eq:SPNPinched} then gives $\delta B_{\mathrm{SPN}}=4.1~\mathrm{fT}/\sqrt{\mathrm{Hz}}$.

\subsection{Fitting of the Measured Field-Response Curves}
We fit the field-response data in Fig.~2 of the main text with the
common PS/SS line shape of Eqs.~\eqref{eq:AmpxPinched} and \eqref{eq:AmpxStretched}, 
\begin{equation}
    y=\frac{A\Omega_0}{\sqrt{\Gamma^4+{(\omega^2-\Omega_0^2)}^2+2\Gamma^2(\omega^2+\Omega_0^2)}}+C.
    \label{eq:fitfunc}
\end{equation}
where $\Omega_0=\Omega_{\rm TS}$ for the PS and
$\Omega_0=\Omega_{\rm L}$ for the SS; $C$ is a common constant
background.

For the PS we take $D_1=-D_0$ and $D_2=D_{-1}=0$, giving
$A_{\rm PS}\propto F(F+1)|D_0|$. For an ideal SS,
$A_{\rm SS}\propto F$, hence
\begin{equation}
    |D_0|=\frac{A_{\mathrm{PS}}}{(F+1)A_{\mathrm{SS}}}.
\end{equation}
All traces are fitted with a shared background, yielding $C=0.091$ mrad/nT.
The fitted parameters are listed in Table~\ref{tab:fitres}.

\begin{figure}[htbp]
    \centering
    \includegraphics[width=0.7\textwidth]{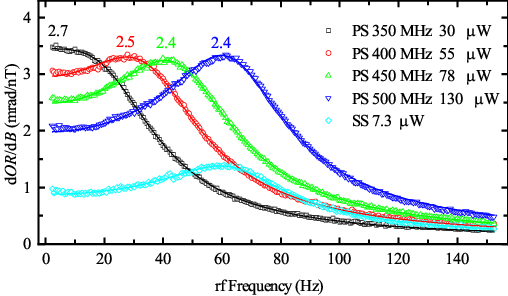}
    \captionsetup{font=small,justification=justified}
    \caption{\label{fitting_results}Fitting results of the field-response curves of Fig.~2 in the main text using Eq.~\eqref{eq:fitfunc}.}
\end{figure}

\begin{table}[htbp]
    \centering
    \captionsetup{font=small,justification=justified}
    \footnotesize
    \caption{\label{tab:fitres}Best-fit parameters of Eq.~\eqref{eq:fitfunc} for the PS and SS data sets. $\Omega_0$ denotes $\Omega_{\mathrm{TS}}$ for PS and $\Omega_\mathrm{L}$ for SS. The enhancement is evaluated as $A_{\mathrm{PS}}/A_{\mathrm{SS}}$. The uncertainties in the enhancement and $D_0$ are obtained from the covariance matrix of the joint fit, which accounts for the correlations among the fitted parameters.}
    \begin{tabular}{lccccccc}
        \toprule
        State & Detuning (MHz) & Power ($\mu$W) & $A$ ($\mathrm{s^{-1}mrad/nT}$) & $\Omega_0/2\pi$ (Hz) & $\Gamma/2\pi$ (Hz) & Enhancement & $D_0$ \\
        \midrule
        PS & 350 & 30  & $5860(40)$ & 21.24(6)  & 22.1(1) & $2.680(20)$ & $0.893(7)$ \\
        PS & 400 & 55  & $5469(18)$ & 35.01(4)  & 21.66(8) & $2.500(14)$ & $0.834(5)$ \\
        PS & 450 & 78  & $5235(14)$ & 45.75(4)  & 21.06(7) & $2.394(12)$ & $0.798(4)$ \\
        PS & 500 & 130 & $5192(11)$ & 63.13(4)  & 20.32(6) & $2.374(12)$ & $0.791(4)$ \\
        \midrule
        SS & 0 & 7.3 & $2187(12)$ & 63.1(10) & 21.5(15) & $1$ (by definition) & $D_2=1 $  (assumed) \\
        \bottomrule
    \end{tabular}
\end{table}

The fitted linewidths, $\Gamma/2\pi=20.3$--$22.1~\mathrm{Hz}$,
are nearly identical for the PS and SS. As expected,
$\Omega_{\rm TS}$ increases with direct-pump power and blue detuning,
while $\Omega_{\rm L}$ is fixed by the bias field. 
The extracted $|D_0|$ decreases from 0.89 to 0.79 as the direct-pump
detuning increases from 350 to 500 MHz. Resolving the tensor splitting
requires $\Omega_{\rm TS}>\Gamma$ and hence substantial blue detuning,
which increasingly excites the competing $F_a\rightarrow F_b'$ transition
and transfers population out of $m=0$. The resulting imperfect
preparation reduces the enhancement below the ideal $F+1=3$. 
Equation (34) neglects $D_2$ and $D_{-1}$ because no resolvable
$3\Omega_{\rm TS}$ feature is observed. Varying $D_2$ over its
physically allowed range, subject to
$D_1=-D_0$ and $D_2=-D_{-1}$, in simulations based on Eq.~(13) changes the recovered $|D_0|$ by less than 0.003, well below the uncertainties in
Table~\ref{tab:fitres}. 

\subsection{Crosstalk between the Two Magnetometer Channels}
\begin{figure}[htbp]
    \includegraphics[width=0.95\textwidth]{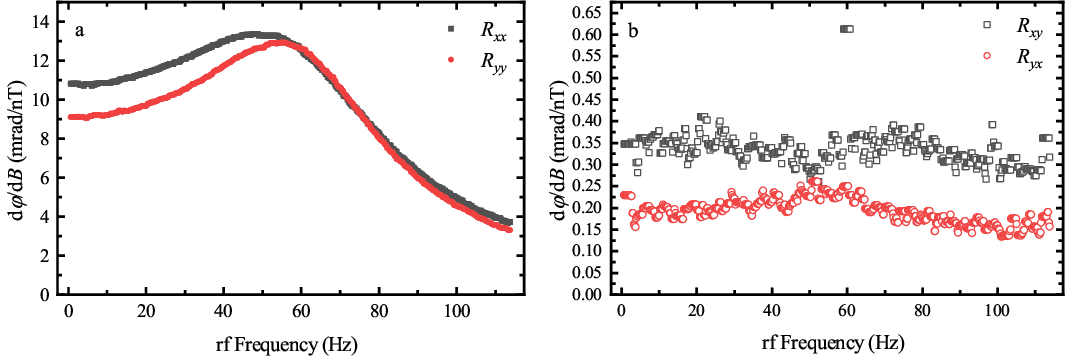}
    \captionsetup{font=small,justification=justified}
    \caption{\label{fig:crosstalk}Magnetometer response matrix under the conditions of Fig.~4 of the main text. (a): Direct responses $R_{xx}$ and $R_{yy}$. (b): Cross responses $R_{xy}$ and $R_{yx}$.}
\end{figure}

Fig.~\ref{fig:crosstalk} shows the channel crosstalk. Defining
$R_{ij}\equiv\partial\phi_i/\partial B_j$, the measured cross responses are of order $2\%$ of the direct responses near resonance.

\FloatBarrier

\bibliography{reference.bib}